\documentclass[
 twocolumn,
 english,superscriptaddress,unsortedaddress,showpacs,showkeys,secnumarabic,nobibnotes,final,prl
 ]
 {revtex4}

\usepackage{graphicx}
\usepackage{amssymb}
\usepackage{color}

\begin{document}

\title{Supermassive Black Hole in an Elliptical Galaxy:\\
Accretion of a Hot Gas with a Low but Finite Angular Momentum }

\author{N.A. Inogamov}

 \affiliation{Landau Institute for Theoretical Physics, RAS, Chernogolovka, 142432, Russian Federation}

 \affiliation{Max-Planck Institut fuer Astrophysik, Garching, D-85741, Germany}

 \author{R.A. Sunyaev}

 \affiliation{Max-Planck Institut fuer Astrophysik, Garching, D-85741, Germany}

 \affiliation{Space Research Institute, RAS, Moscow, 117997, Russian Federation}

\begin{abstract}
 The accretion of hot slowly rotating gas onto a supermassive black hole is considered.
 The important case where the velocities of turbulent pulsations at the Bondi radius $r_B$ are low,
   compared to the speed of sound $c_s,$ is studied.
 Turbulence is probably responsible for the appearance of random average rotation.
 Although the angular momentum at $r_B$ is low, it gives rise to the centrifugal barrier
   at a depth $r_c = l^2/G M_{BH} \ll r_B,$ that hinders supersonic accretion.
 The numerical solution of the problem of hot gas accretion with finite angular momentum is found
   taking into account electron thermal conductivity and bremsstrahlung energy losses
     of two temperature plasma for density and temperature near Bondi radius
       similar to those observed in M87 galaxy.
 The saturation of the Spitzer thermal conductivity was also taken into account.
 The parameters of the saturated electron thermal conductivity were chosen similar to the parameters
   used in the numerical simulations of interaction of the strong laser beam radiation with plasma targets.
 These parameters are confirmed in the experiments.
 It is shown that joint action of electron thermal conductivity and free-free radiation
   leads to the effective cooling of accreting plasma and formation of the subsonic settling of accreting gas
     above the zone of a centrifugal barrier.
 A toroidal condensation and a hollow funnel that separates the torus from the black hole
   emerge near the barrier.
 The barrier divides the flow into two regions: (1) the settling zone with slow subKeplerian rotation
   and (2) the zone with rapid supersonic nearly Keplerian rotation.
 Existence of the centrifugal barrier leads to significant decrease of the accretion rate $\dot M$
   in comparison with the critical Bondi solution for $\gamma=5/3$
     for the same values of density and temperature of the hot gas near Bondi radius.
 Shear instabilities in the torus and related friction cause the gas to spread slowly
   along spirals in the equatorial plane in two directions.
 As a result, outer $(r>r_c)$ and inner $(r<r_c)$ disks are formed.
 The gas enters the immediate neighborhood of the black hole or the zone of the internal ADAF flow
   along the accretion disk $(r<r_c).$
 Since the angular momentum is conserved, the outer disk removes outward an excess of angular momentum
   along with part of the matter falling into the torus.
 It is possible, that such outer Keplerian disk was observed by Hubble Space Telescope
   around the nucleus of the M87 galaxy in the optical emission lines.
 We discuss shortly the characteristic times during which the accretion of the gas with developed turbulence
   should lead to the changes in the orientation of the torus, accretion disk and, possibly, of the jet.
\end{abstract}

 %  *** *** *** pacs



\maketitle

 Full text of the paper with two Appendices is published in Astronomy Letters,
  Vol. 36, p. 835 (2010).

 \section{INTRODUCTION}

 The nearest giant elliptical galaxy M87 hosts a supermassive black hole (SMBH)
   with a mass of $(3-6) \times 10^9 M_{Sun}$
     (Ford et al. 1994; Macchetto et al. 1997; Gebhardt and Thomas 2009)
       and is filled with a hot gas that intensely radiates in X-rays.
 The ratio of the gravitational radius $r_g =2 G M_{BH}/c^2\approx 10^{15}$cm
   to the distance of M87 from the Earth 17.9 Mpc (Gebhardt and Thomas 2009)
     is known to correspond to an angular size of 4-8 $\mu\!$as
       and to be only slightly inferior in this parameter
         to the black hole (BH) SGR $A^*$ at the center of our Galaxy.
 In contrast to SGR $A^*,$ the nucleus of M87 is observable in the optical and ultraviolet bands
   and is actively investigated in all ranges of the electromagnetic spectrum.
 The X-ray luminosity $L_X \sim 7 \times 10^{40}$erg$\cdot$s$\!^{-1}$ (Di Matteo et al. 2003)
   of the central 100 pc in M87 is negligible compared to the Eddington one.
 At the same time, the jet in M87 is observed in the radio, optical, and X-ray bands,
   while the absence of a "cooling flow" in the hot gas of the central regions of M87
     is indicative of its heating by the "mechanical" energy of the jet, shocks, and gas outflows
       maintained by BH activity.

 There are different opinions on the nature of this activity - in principle,
   it can be related to BH spin-down (Blandford and Znajek 1977),
     but the closeness (in order of magnitude) of the "mechanical" luminosity
       $L_m\sim 3\times 10^{43}$erg$\cdot$s$\!^{-1}$ (Churazov et al. 2002; Di Matteo et al. 2003)
         and the power of the accretion energy release
           expected in the picture of spherically symmetric Bondi accretion
             assuming the accretion efficiency to be $\sim$10\% of $\dot M_B c^2$
               (Owen et al. 2000; Young et al. 2002; Di Matteo et al. 2003) is suspicious.
 The point is that the Chandra X-ray telescope resolves the zone of the Bondi radius $r_B \sim 10^5 r_g,$
   but the complex structure of the brightness distribution
     allows the temperature and electron density of the fully ionized gas
       to be determined only on average over a zone of $15''-20''$
         whose size is more than an order of magnitude larger than the Bondi radius
           (Forman et al. 2007).
 The temperature and density along with the SMBH mass determine the Bondi accretion rate $\dot M_B.$

 It is the assumption about such energy release during spherically symmetric Bondi accretion
   that is commonly made now in cosmological simulations of feedback
     from rapidly growing BHs in the nuclei of forming or merging galaxies
       on their growth and the outflow of matter in these galaxies (Springel et al. 2005).
 This feedback is known to be very important for the evolution of galaxies.

 In this paper, we would like to recall that a low but finite angular momentum of the accreting gas in galaxies
   plays a very important role.
 Since the ratio $r_B/r_g \sim (c/c_s)^2 \sim 10^5$ is very large,
   the presence of tangential gas velocities
     $v_{\varphi} \sim c_s(c_s/c) \sim 3 \times 10^{-3}c_s \sim 2$km s$\!^{-1}$ in the zone of the Bondi radius
       should lead to the appearance of a centrifugal barrier at $r_c >r_g$
         and the stopping of a spherically symmetric flow
           (see, e.g., Kolykhalov and Sunyaev 1980,
             see also Limber, 1964; Fabian, Rees, 1995; Beskin and Malyshkin, 1996).
 In what follows, $c_s$ is the speed of sound; $c_s \sim 700$km s$\!^{-1}$ at the Bondi radius.

 M87 has an anomalously low (even for elliptical galaxies) "spin"
   that characterizes the rotation of the galaxy's stellar component.
 SAURON data (Emsellem et al. 2007) give only an upper limit for the spin,
   which, nevertheless, corresponds to rotation velocities much higher than 2 km s$\!^{-1}.$
 The gas in the zone of the Bondi radius is unlikely to have fallen there as a result of the mass loss by stars.
 However, the hot gas near the Bondi radius is most likely subjected to turbulent motions
   that arise from the mergers of galaxies with M87 and the activity of its nucleus
     (for a discussion, see Appendix 1).
 It is these motions that impart a finite angular momentum to the gas
   that determines the flow pattern in the zone $r \sim r_c \ll r_B.$

 In the absence of mechanisms for efficient cooling of the accreting gas at $r \sim (10-100)r_g,$
   advection-dominated accretion flows (ADAFs) can emerge in principle below $r_c.$
 In addition, the angular momentum should somehow be withdrawn beyond $r_B$ upward - against the gas flow.
 ADAF-type solutions have been suggested long ago (Narayan and Yi 1994; Fabian and Rees 1995).
 They are actively discussed as applied to both SGR $A^*$ and M87 (Fabian and Rees 1995)
   and are distinguished by a low radiative cooling efficiency during accretion.
 Below, we will show that another regime of accretion works in elliptical galaxies similar to M87.

 The combined effects of "saturated" electron thermal conduction,
   deviation of the electron temperature from the ion one $T_e <T_i,$
     and bremsstrahlung near $r_c$ are capable of cooling the accreting plasma flow
       and contributing to the establishment of delayed gas settling into the comparatively cold torus
         near the location of the centrifugal barrier.
 In the presence of magneto-rotational instability (Velikhov 1959; Balbus and Hawley 1991),
   the torus should spread out in both directions along the radius and form a standard accretion disk
     that carries matter to the BH
       and an outer accretion disk at $r>r_c$
         responsible for the removal of angular momentum into the zone beyond the Bondi radius
           (see Kolykhalov and Sunyaev 1980).
 Numerical simulations show that the accretion rate for such a solution
   is only an order of magnitude lower
     than that in the critical regime of Bondi accretion for a gas with the adiabatic index $\gamma =5/3$
       and depends on the initial angular momentum,
         i.e., on the position of the centrifugal barrier.
 As the angular momentum grows and $x_c = r_c/r_B$ changes from 0.01 to 0.04,
   the accretion rate in the considered mode increases from 8\% to 17\%
     of the accretion rate in the critical Bondi flow with $\gamma =5/3$ (see the Figure at the end of paper).

 Our interest in this problem was heightened noticeably by the assumption that the outer disk
   above the centrifugal barrier predicted by Kolykhalov and Sunyaev (1980)
     could be associated with the Keplerian disk observed by the Hubble Space Telescope
       by emission in the H$\!\alpha,$ [O II], [N II], and other lines
         (Ford et al. 1994; Harms et al. 1994; Macchetto et al. 1997)
           at distances $0.03 <r/r_B < 1$ from the BH.
 It is this disk that allowed the authors of the papers cited above
   to determine the BH mass in M87 just as was subsequently done for the BH in NGC 4258
     from maser radio lines (Miyoshi et al. 1995).

 The suggested picture of accretion has quite a few observational consequences being discussed here
   and outstanding points on which we continue to work.
 First of all, the modification of the solution by Kolykhalov and Sunyaev (1980) for the case of a disk
   that exists due to the removal of angular momentum
     and that is under the external pressure of the atmosphere of hot gas
       settling from the zone near the Bondi radius to the centrifugal barrier is of considerable interest.
 Allowance for the heating of this disk outside by the electron thermal conduction of the hot gas
   is also important.
 It is this mechanism that is most likely responsible for the disk heating
   and that provides its comparatively high luminosity in optical lines.

 The most important observational consequences to be discussed below are the following:

 (1) An enhanced brightness in X-rays and ultraviolet lines of the narrow zone
   of intense gas cooling above the torus near the centrifugal radius
     where $\sim 10^{-4}-10^{-3}$ of the total mechanical luminosity
       of the M87 nucleus should be radiated away.
 This luminosity should not exceed the X-ray luminosity of the central source,
   which (according to a possible overestimate) reaches $L_X \sim 7 \times 10^{40}$erg s$\!^{-1}$
     (Di Matteo et al. 2003).

 (2) The influence of turbulent motions in the zone with $r>r_B$ on the characteristic time
   in which the jet direction changes (see Appendix 1).

 (3) The emission from the disk above the centrifugal radius in ultraviolet lines
   through its heating by electron thermal conduction from the hot accreting atmosphere.

 (4) The effects from the eclipse of the central part of the counterjet by the disks being discussed.

 Unfortunately, the suggested picture cannot explain the high ratio
   of the mechanical luminosity of the M87 nucleus to its bolometric luminosity
     in all ranges of the electromagnetic spectrum.
 At the very least, this requires assuming that the solution of a thin standard disk
   (Shakura and Sunyaev 1973) ceases to work in the zone with $r< (20-100)r_g$
     and transforms into ADAF.

 \section{OVERALL PICTURE}

 \subsection{Spatial Structure}

 The flow scheme is presented in Fig. 1.
 The Bondi sphere of radius $r_B$ bounds the zone of SMBH gravitational influence.
 Outside it, the temperature and density asymptotically reach their values at infinity,
   $T_\infty$ and $n_\infty.$
 The subsonic atmosphere with the settling of matter along the streamlines
   indicated by the dashed curves in Fig. 1
 is located between the spheres with radii $r_B$ and $r_c:$
   \begin{equation}
   r_c = l^2 /G M_{BH} = ({\rm Ma}_\varphi|_B)^2 r_B = x_c r_B,
   \end{equation}
 where ${\rm Ma}_\varphi|_B$ is the Mach number relative to rotational velocity at radius $r_B.$
 The term "atmosphere" is justified, because, as we show here,
   the barrier reduces greatly the rate of accretion $\dot M.$
 As a result, the latter is realized in the regime of slow settling
   with radial velocities of the order of a few percent of the local speed of sound.
 Accordingly, the inertial effects and heat release through braking in radial velocity here
   are unimportant, while the rate $\dot M$ is appreciably lower
     than the maximum possible one in a transonic (at the adiabatic index $\gamma = 5/3),$
       spherically symmetric Bondi flow.
 Note that such a transonic flow is often called critical.
 At $\gamma = 5/3,$ the critical flow reaches the speed of sound at point $r=0$
   (Stanyukovich 1960; Kato et al. 2008) and, hence, is transonic.

% ------- -- --- --- --- --- --- --- ------- Ris-01
\begin{figure}[t]
 \includegraphics[width=1\columnwidth]{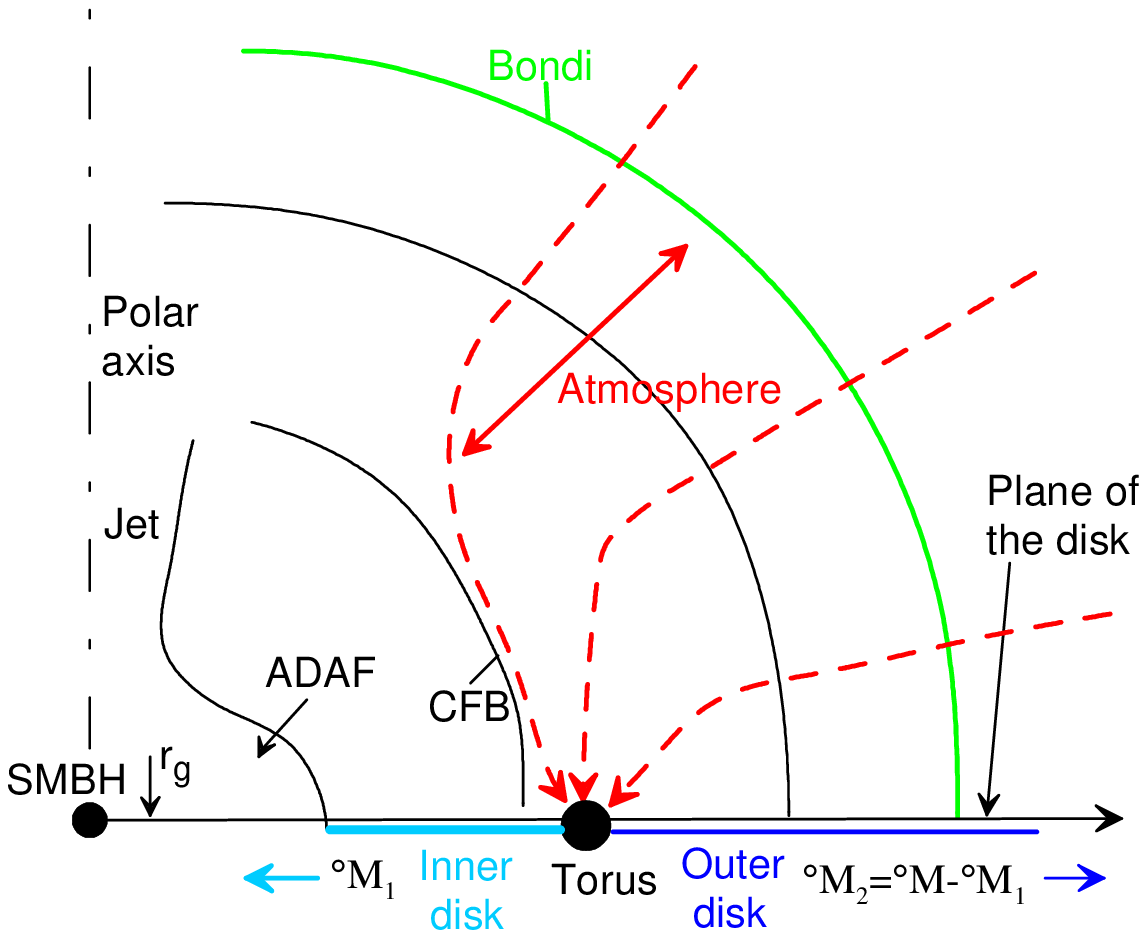}
 \caption{\label{fig:1}
 Structure of the accretion flow in the presence of a low angular momentum.
 Rotation causes the flow
   (that is close to spherically symmetric settling above the centrifugal barrier $r>r_c)$
     to be restructured qualitatively.
 Restructuring affects not only the regions near and below the barrier $r<r_c$
   but also above it $r>r_c.$
 More specifically, a dense outer disk that threads the rarefied atmosphere
   and that transports the matter back, i.e., toward the settling flow, appears.
      }
\end{figure}

 The scale $r_c$ (1) defines the location of the centrifugal barrier (CFB) and the centrifugal funnel;
   in what follows, $l$ is the specific angular momentum.
 This scale is small $(r_c \ll r_B)$ if the Mach number ${\rm Ma}_\varphi|_B = (v_\varphi|_B)/(c_s|_B)$
   in rotation velocity at the Bondi radius is small.
 A typical value of ${\rm Ma}_\varphi|_B \sim 0.1$ is determined by subsonic turbulence
   beyond the radius $r_B$ (see Appendix 1).
 In the case of low ${\rm Ma}_\varphi|_B,$ the atmosphere occupies the geometrically extended segment $r_c <r<r_B.$

 Above the CFB in Fig. 1, the dynamical influence from the presence of angular momentum in the settling flow
   is weak - the centrifugal force can then be neglected.
 Therefore, at $r>r_c,$ the flow is radial with approximately straight streamlines.
 At $r \sim r_c,$ rapid rotation that accounts for a significant fraction of the Keplerian one
   begins to dominate.
 The flow is blocked by the CFB "wall" near which the streamlines curve greatly (see Fig. 1).
 The minimum of the gravitational and centrifugal potentials, $\Phi_{sum} = -GM/r + l^2/2r^2_{cyl},$
   forms the torus center in Fig. 1; in what follows, $r_{cyl}$ is the cylindrical radius
     equal to the distance to the polar axis.
 The subsonic accretion flow slowly sinks toward the minimum of the potential $\Phi_{sum}$ (see Appendix 2).

 The trajectories of individual particles are spirals running over the surface
   that is formed by rotating the dashed curve (streamline) in Fig. 1 around the polar axis.
 The spiral pitch $h_{sp}$ in the meridional plane shown in Fig. 1 is equal to the distance
   traversed by a particle in one turn around the axis.
 It is defined by the velocity ratio $v_\varphi/\sqrt{v^2_r + v^2_{\theta}},$
   where $v_\theta$ is the meridional velocity.
 The pitch $h_{sp}$ decreases rapidly as the minimum of $\Phi_{sum}$ is approached,
   because the rotation velocity begins to exceed greatly the settling velocity.
 This is attributable to strong cooling and a sharp density rise toward the minimum.
 The ratio $h_{sp}/r$ (where $r$ is the local radius) is great at $r>r_B,$
   of the order of unity in the atmosphere at ${\rm Ma}_\varphi|_B \sim 0.1,$
     and, as was said above, small near the barrier.

 A simplified scheme with a list of the structure's main links, flow separation at the torus center,
   and an hierarchy of scales is presented in Figs. 2 and 3.
 For clarity, Fig. 2 is not to scale: the region of the barrier $r_c$ is greatly magnified.
 Figure 3 shows the actual ratio of the sizes.
 In Fig. 3, the turbulent velocity field described in Appendix 1 occupies the region
   from an outer size of $2-5$ kpc to the Bondi sphere $r_B;$
     the Bondi sphere $r_B \sim 10^5 r_g$ is located inside the turbulent field;
       the CFB $r_c \sim 10^3 r_g$ under the Bondi sphere is two orders of magnitude deeper
         (the typical condition ${\rm Ma}_\varphi|_B \sim 0.1$ is assumed to be met);
           the thin disk - ADAF transition (point $r_A$ in Fig. 3)
             occurs somewhere within the centrifugal radius $r_c;$
               the radius of the last stable orbit around the SMBH is $3 r_g;$
                 the jet base lies in a region of $\sim 10 r_g.$

% ------- -- --- --- --- --- --- --- ------- Ris-02
\begin{figure}[t]
 \includegraphics[width=1\columnwidth]{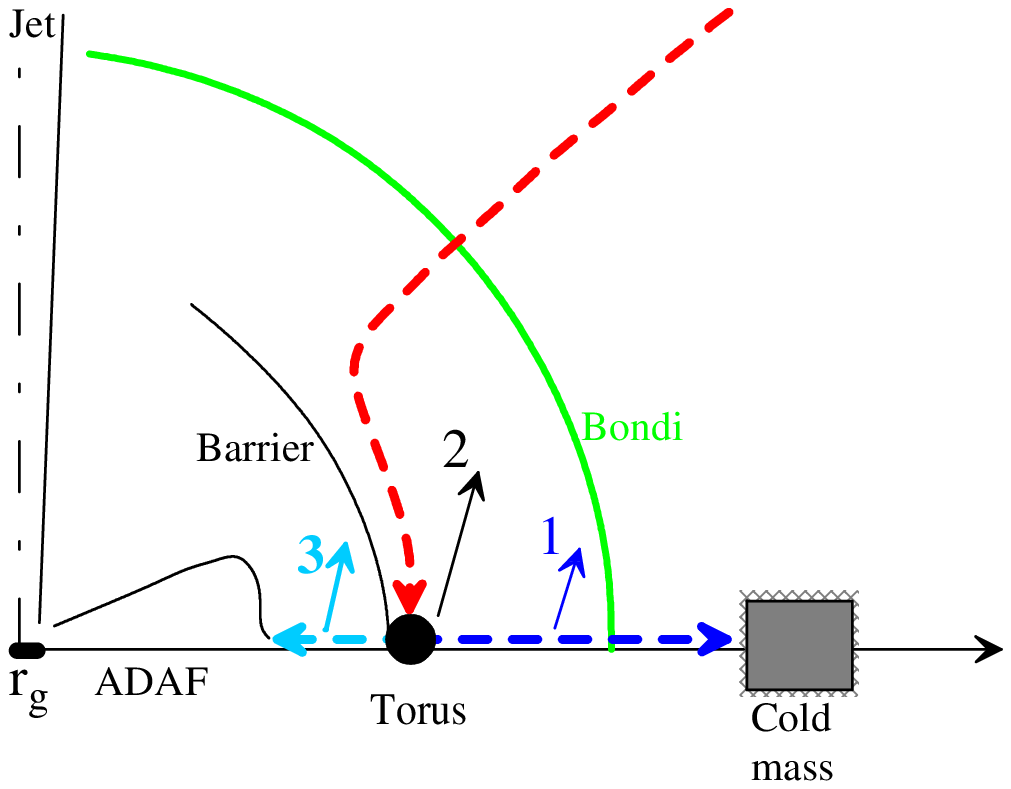}
 \caption{\label{fig:2}
 Scheme of physical processes: 1 - ultraviolet emission from the outer disk removing angular momentum,
  2 - torus cooling by X-ray emission, 3 - optical thin-disk emission.
 The ADAF contribution to the emission is very small.
      }
\end{figure}

% ------- -- --- --- --- --- --- --- ------- Ris-03
\begin{figure}[t]
 \includegraphics[width=1\columnwidth]{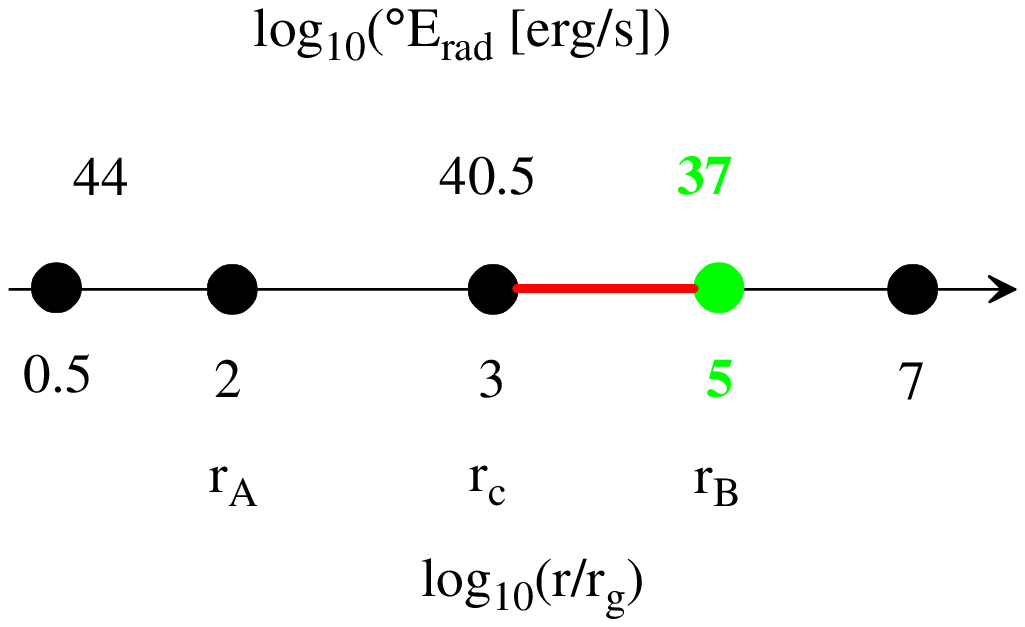}
 \caption{\label{fig:3}
 Main scales: turbulence covers the outer segment from 2-5 kpc to $r_B;$
   the Bondi radius is $r_B \sim 10^5 rg;$
     $r_c \sim 10^3 r_g$ is the barrier position; $r_A$ is the thin disk-ADAF transition;
       the radius of the last stable orbit around the SMBH is $3 r_g.$
      }
\end{figure}

 In Fig. 3, the X-ray luminosity of the hot gas from a volume equal to the volume of the Bondi sphere
   is approximately three orders of magnitude lower than that from the barrier neighborhood.
 This, in turn, is three to four orders of magnitude lower than the mechanical luminosity.
 An energy of such an order of magnitude, first, is carried away by advection in the SMBH
   and, second, is transferred mechanically by the relativistic jet into the surrounding space.
 The thin disk - ADAF transition (Narayan and Yi 1994) is caused by dissipative heat release
   through viscous friction.
 The plasma does not cope with the emission of energy from this heat release due to a small density,
   because $\dot M \sim 10^{-4} \dot M_{Edd}$ under our conditions,
     where $\dot M_{Edd}$ is calculated from $\dot E_{Edd}$ with a canonical coefficient of 0.1.
 In the ADAF regime, the flow thickness above the equatorial plane is of the order of the radius,
   as is shown in Fig. 2, while the coefficient of $\alpha$-friction is great: $\alpha \sim 1.$
 The boundary of the disk - ADAF transition $r = r_A$ could be established
   if the energetically most weighty contribution from emission 3 in Fig. 3,
     which, besides, differs spectrally from contributions 1 and 2,
       were detected and identified in observations.

 The centrifugal torus that "props up" the atmosphere from below
   is thick under weak cooling of the atmosphere and neighborhoods of the torus,
     with a "doughnut" radius $r_{tor}$ of the order of $r_c.$
 As will be seen below, such is the situation in the absence of electron thermal conduction,
   because the cooling time $t_{Brs} \sim 10^8$yr through hot-plasma bremsstrahlung near $r_B$
     is comparable to the characteristic times for the cooling flows in the central regions of galaxies
       and clusters.
 It is very long for the accretion problem under consideration,
   where the Keplerian period $t_K|_B$ at the Bondi radius $r_B$ is $\sim 10^5$yr.

 % The accretion essentially ceases under insufficient cooling,
 %  because the rate of angular momentum removal from a thick torus to infinity through an extended atmosphere
 %    is low.
 In the paper it is supposed that the stable stratification of the atmosphere in the SMBH gravity field
   prevents angular momentum transport through the atmosphere.
 Therefore this transport is going through the outer disk.
 The stable stratification is formed due to cooling of the bottom of the atmosphere near the torus.
 The cooling causes the entropy drop
 \begin{equation}
   ds/dr >0
   \end{equation}
 with decrease of radius in the atmosphere.

 % slow differential rotation in the atmosphere,
 %  because the rotation velocity $v_\varphi$ in the atmosphere is subsonic
 %    $v_\varphi \ll v_K,$ $v_\varphi \ll c_s,$ where $v_K(r)$ is the local Keplerian velocity
 %      and $c_s(r)$ is the speed of sound.
 % Consequently, the shear frequency $v_{shear} = dv_\varphi/dr = l/r^2$ is low
 %  compared to the local Keplerian frequency $\nu_K.$
 % Accordingly, the mechanical force coupling between differentially rotating atmospheric layers is weak.
 % The force interaction between layers is known to arise from the forces of viscous friction
 %  produced by turbulence and magnetic fields $(\alpha$-viscosity).
 % In this case, the tangential stress $w_{r\varphi}$ is proportional to the rotation velocity gradient:
 % \begin{equation}
 %  w_{r\varphi} = \rho \nu_{turb} \nu_{shear},
 %  \end{equation}
 %   where $\nu_{turb}$ is the effective kinematic viscosity.

 The detection of emission 1 in Fig. 2 allowed the presence of a thin disk in M87 to be firmly established
   (Harms et al. 1994; Ford et al. 1994; Macchetto et al. 1997).
 In our model with an outer disk in Fig. 1, this emission is associated with the transition layer
   between a cold disk and a hot overlying atmosphere.
 The expenditure on emission is covered through transition-layer heating by an electron heat flow
   going from the atmosphere.
   % and, possibly, through friction between a highly dense disk rotat- ing with the Keplerian velocity
 % and a barely moving rare?ed atmosphere (supersonic friction). In the latter case, a force that gradually brakes
 % the outer disk layers appears. Since the velocity di?erence at the transition-layer boundaries is much greater
 % than the speed of sound in the disk, it is natural to call this braking supersonic friction.

 \subsection{Torus Cooling and Angular Momentum Removal
over the Region above $r_c$ through a Barely Moving Atmosphere}

 The picture changes qualitatively from the case with virial atmosphere
   in the presence of efficient torus cooling
     through electron thermal conduction and radiative heat transfer.
 The torus thickness $r_{tor}$ decreases greatly $(r_{tor} \ll r_c)$
   and the doughnut becomes thin (see Fig. 1).
 As has been said in the preceding section, the matter sinks into the thin torus along the streamlines
   indicated by the dashed curves in Fig. 1 as it cools down.
 The matter in the torus is in rapid centrifugal rotation.
 Since the torus temperature is considerably lower than the virial temperature $T_{vir}(r_c),$
   the rotation velocity $v_\varphi(r_c) \approx v_K(r_c)$ exceeds significantly the speed of sound $c_s(r_c).$

 Under conditions of supersonic centrifugal rotation, the a friction becomes dynamically efficient.
 Due to friction, dense outer and inner disks of small thickness $h \ll r$ grow from the thin dense torus -
   the thin doughnut is spread into a disk (see Fig. 1).
 This is caused by strong mechanical coupling of differentially rotating rings
   with different radii inside the disk through friction.
 The disks are dense, because bremsstrahlung, which became efficient
   when the temperature dropped and the density rose exponentially in the torus,
     remains decisive in the disks as well.
 In the disks, the shear frequency $dv_\varphi/dr = \nu_{shear} \approx \nu_K$
   and, hence, the friction stress $w_{r\varphi}$ is significant.
 As has been said above, for M87 the inner disk is restructured into ADAF at some distance $r_A$
   within the radius $r_c$ due to a low cooling efficiency by radiation in this disk.

 In the torus, the downward-settling flow separates into two flows:
   one goes rightward into the outer disk,
     while the other goes leftward into the inner disk (see Fig. 2).
 Thus, the ring of feeding of these disks is associated with the torus,
   as was assumed in the Kolykhalov–Sunyaev model (Kolykhalov and Sunyaev 1980).
 Through the atmosphere (toward the accretion flow!), the outer disk removes,
   first, the angular momentum of the matter settling through the atmosphere
     and, second, the angular momentum of the inner-disk matter and the ADAF,
       while the inner disk, which is a standard one in the segment $r_A <r< r_c$ (Shakura and Sunyaev 1973),
         supplies the ADAF, the jet, and the SMBH with a mass flow.

 \section{SYSTEM OF EQUATIONS}

 Under conditions of the galaxy M87, the Knudsen number at the Bondi radius $r_B$ (estimated from $r_B)$
   is small.
 Near and above $r_B,$ there are enough collisions
   to equalize the electron, $T_e,$ and ion, $T_i,$ temperatures.
 However, for slow rotation at $r_B$ considered here, the CFB $r_c$ (1)
   lies deep under the Bondi radius $r_c \ll r_B,$ $x_c \ll 1,$ $x = r/r_B,$ $x_c = r_c/r_B.$
 Accordingly, there is much heat being released as the matter sinks to this depth.
 The removal of such an amount of heat gives rise to a region of steep temperature gradients
   at a sufficient depth.
 At this depth, the Knudsen number increases, the collisions are not enough,
   the thermal balance between electrons and ions is upset, and $T_e$ becomes lower than $T_i.$
 Therefore, the system of our equations is written for a two-temperature plasma:
 \begin{equation}
 \partial\rho/\partial t + {\rm div}(\rho\vec v)=0,
 \end{equation}
 \begin{equation}
 \rho\partial\vec v/\partial t + \rho(\vec v\cdot\nabla)\vec v+\nabla p +\rho\nabla\Phi=0,
 \end{equation}
 \begin{equation}
 \rho\frac{\partial e_e}{\partial t} + \rho(\vec v\cdot\nabla)e_e+p_e {\rm div}\vec v +{\rm div} \vec q +
 \dot e - \alpha(T_i-T_e)=0,
 \end{equation}
 \begin{equation}
 \rho\partial e_i/\partial t + \rho(\vec v\cdot\nabla)e_i+p_i {\rm div}\vec v + \alpha(T_i-T_e)=0.
 \end{equation}
 The system that consists of the continuity equation (3), the dynamical equation (4),
   and the thermal balances for electrons and ions (5) and (6) describes the flow
     from infinity to the CFB (accretion through the atmosphere).
 In the atmosphere, we neglect the $\alpha$-viscosity in Eqs. (4) and (6).
 The molecular viscosity is low.
 The partial pressures and energies in Eqs. (5) and (6) are
 $$ p_e = n k_B T_e,\;\; e_e =(3/2) k_B T_e/m_p,$$
 $$ p_i = n k_B T_i,\;\; e_i =(3/2) k_B T_i/m_p.$$
 The total pressure and the gravitational potential in Eq. (4) are
   $$ p = p_e + p_i, \;\; \Phi= -G M/r.$$

 The electron heat flux $q$ in balance (5) is significant.
 Let us represent it as a harmonic mean,
 \begin{equation}
  q = \zeta_{ehc} \left(\frac{ 1}{\kappa_{Sp} |\nabla T_e|} +
  \frac{1}{q_{lim}}\right)^{-1} \frac{(-\nabla T_e)}{|\nabla T_e|}.
 \end{equation}
 This form allows the constraint on the flux $q$ to be taken into account
   at high Knudsen numbers (see, e.g., Back et al. 1996).
 Equation (7) is composed of the Spitzer expression
 \begin{equation}
 \kappa_{Sp} = \frac{1.84 \times 10^{-5}}{\ln\Lambda} T^{5/2}_e
  {\rm [erg \; s}^{-1} {\rm K}^{-1} {\rm cm}^{-1} ]
 \end{equation}
  and the limiting flux
   \begin{equation}
 q_{lim} = f n k_B T_e \sqrt{\frac{k_B T_e}{m_e}}
 = f\, 0.3^{3/2} \sqrt{\frac{m_p}{m_e}} \rho c^3_s.
 \end{equation}
 In (8) the temperature $T_e$ is in Kelvins,
   $\ln\Lambda = 30-40$ is the Coulomb logarithm;
     in (9) we took $c_s = \sqrt{(10/3) k_B T_e/m_p.}$
 The coefficient $\zeta_{ehc} =0.1-1$ in (7) allows for the decrease in thermal conductivity
   due to the influence of the magnetic field associated with MHD turbulence
     (see, e.g., Zakamska and Narayan 2003).

 The constraint on the Spitzer thermal conductivity is taken into account
   using Eqs. (7), (8), and (9).
 It becomes necessary if the heat flux $q$ calculated from Spitzer's formula (Spitzer 1962)
  $q_{Sp} = -\kappa_{Sp} T'_{e}$ exceeds the vacuum flux
    $n k_B T_e \sqrt{k_B T_e/m_e},$ where $m_e$ is the electron mass.
 This occurs in the region of steep temperature gradients
   when the characteristic length $T_e/|T'_e|$ becomes comparable to the electron mean free path
     or the mean free path becomes of the order of the radius.
 The harmonic mean (7) is used to constrain the flux in the region of steep gradients.
 Due to the development of plasma instabilities, the limiting parameter $f$ turns out to be small
   (Back et al. 1996).
 We will vary it within the commonly assumed range $f =0.05-0.3.$
 At $f =0.05,$ the combination $f\, 0.3^{3/2} \sqrt{m_p/m_e}$ in (9) is 0.35
   and, accordingly, $q_{lim} =0.35 \rho c^3_s.$

 The radiative losses were calculated from the formulas
 $$ \dot e = \dot e_{brs} + \dot e_{compt}, $$
   \begin{equation}
   \dot e_{brs} =2.1 \times 10^{-27} n^2 \sqrt{T_e},
   \end{equation}
   \begin{equation}
 \dot e_{compt} = 4 \sigma_T k_B (T_e -\langle T_{rad}\rangle )\varepsilon_{rad} n_e /m_e c,
   \end{equation}
   $$
 \varepsilon_{rad} = L/4\pi r^2 c,
   $$
 where $\dot e_{brs}$ and $\dot e_{compt}$ are given in ${\rm [erg} \; {\rm s}^{-1} {\rm cm}^{-3} ],$
   $n$ and $n_e$ are in $[{\rm cm}^{-3}]$ and $\eta=0.1.$
 Equation (5) is the energy balance between the change in electron internal energy
   in a Lagrangian particle,
     (i) the work for adiabatic expansion of matter $p_e {\rm div}\vec v$
         under the electron pressure,
     (ii) the heat transfer by thermal conduction (7),
     (iii) the cooling through bremsstrahlung losses (10),
     and (iv) the Compton cooling (at $T_e > 4 \langle T_{rad}\rangle)$
       due to the scattering of photons from the central X-ray source by electrons (11).

 This central source is associated with the radiation from a region close to the BH.
 The role of electron-photon heat exchange in quasars and in Seyfert galactic nuclei
   is discussed in Levich and Sunyaev (1971).
 The formula for Compton energy exchange between electrons and photons (11)
   includes the difference between the electron and photon temperatures;
     $\langle T_{rad}\rangle \sim 1$keV is the effective temperature of the photon gas,
       electrical neutrality $Z n_e = n_i = n$ is assumed, the plasma is a fully ionized,
         hydrogen one $Z=1,$ $\sigma_T = 6.652 \times 10^{-25}\;{\rm cm}^2$
           is the Thomson cross section,
             $\varepsilon_{rad}$ is the radiation energy density from the central source,
               $L$ is the luminosity of this source.

 The expressions for electron-ion heat exchange are
   \begin{equation}
  \alpha (T_i - T_e)=3 (m_e/ m_p) \nu n k_B (T_i - T_e),
   \end{equation}
 $$
 \nu = (4/3)\sqrt{2\pi} Z n_e e^4 \ln\Lambda/m_e^{1/2} (k_B T_e)^{3/2}.
 $$
  As has been said above, we assume that $Z =1.$
  In (12) $\nu$ is the electron-ion collision frequency (Spitzer 1962).

 \section{SPHERICALLY SYMMETRIC
APPROXIMATION OF A FLOW IN AN ATMOSPHERE}

 This Section is devoted to derivation of non-dimensional steady-state equations
   from dimensional equations presented above.
 The system of non-dimensional steady-state equations is used for numerical integration.
 We omit this technical Section in the text for astro-ph.
 The full paper is published in Astronomy Letters, 2010, Vol. 36, No. 12, pp. 835-847.

 \section{ACCRETION ATMOSPHERE IN THE GALAXY M87}

 The boundary conditions that complement system (14)-(16) from previous Section are
 $$
 \hat n(x = \infty)=1,\; t_e(\infty)=1,\; t_i(\infty)=1,\; t'_e(x_c)=0, \eqno (17)
 $$
 where normalized quantities $\hat n=n/n_\infty,$ $t_e=T_e/T_\infty,$ $t_i=T_i/T_\infty$
    are used.
 The last of conditions (17) follows from the requirement that there be no electron heat exchange
   between the torus and the disk (see Figs. 1 and 2).

 The unknown accretion rate $\dot M$ (13) enters into the fourth-order system (14)-(16)
   with four conditions (17) as a parameter.
 Therefore, the solution $(\hat n, t_e, t_i)$ will depend on the non-dimensional parameter $\dot m:$
   $$ \hat n(x;\dot m), \;\, t_e(x;\dot m), \;\, t_i(x;\dot m), $$
 where
 $$
 x = \frac{r}{r_B}, \; \;\; \dot m = \frac{\dot M}{\dot M_{max}(\gamma=5/3)},
 $$$$
 \dot M_{max}\left( \gamma = \frac{5}{3} \right)
  = \frac{1}{4} \dot M_B\left(\gamma = \frac{5}{3}\right),
  $$
  % $$
  % \hat v = \frac{v}{c_{T\infty}},\; c_{T\infty} = \sqrt{\frac{2 k_B T_\infty}{m_p}},
  %$$
  $$
 \dot M_B(\gamma) = 4\pi [r_B(\gamma)]^2 \rho_\infty c_\infty(\gamma),
 $$$$
  r_B = \frac{G M}{c_\infty(\gamma)^2}, \; \;\;
  c_\infty(\gamma)= \sqrt{\frac{2 \gamma k_B T_\infty}{m_p}}.
 $$

 The set of such solutions forms a one-parameter family (the $\dot m$ runs through the family).
 From this family we will select the unique solution (the unique value of $\dot m)$
   that corresponds to the maximum of the ratio
   $$
    (T_{vir}|_c)/(T_e|_c) \eqno (18)
   $$
 of the virial temperature $T_{vir}(r)$ at the CFB $r_c, x_c$
   to the electron temperature in the atmosphere near the barrier.

 The minimum of the density ratio $(\rho_{vir}|_c)/(\rho|_c)$ (see Appendix 2)
   corresponds to the largest temperature ratio $(T_{vir}|_c)/(T_e|_c)$ (18).
 In an adiabatic atmosphere without any losses
   through electron thermal conduction (15) and radiative cooling (10),
     the virial density distribution $\rho_{vir}(r)$ is formed.
 The solution of system (14)-(17) that includes the thermal losses
   is given by the distribution $\rho(r).$
 The torus is thinnest for the highest density (see Figs. 1 and 2).
 The inner and outer disks are then thinnest (among the solutions of the one-parameter family).
 At a large ratio of the densities of the outer disk and the atmosphere,
   the outer disk can exist for a long distances inside the hot rarefied atmosphere,
     despite the great difference in temperatures and rotation velocities.
 In this case, the dense outer disk threads the atmosphere and is retained up to the Bondi radius
   or even farther (Fig. 2).
 Thus, the angular momentum of the settling matter is removed outward.

 A typical example of integrating system (14)-(17) with the solution selection condition (18)
   is shown in Figs. 4 and 5.
 This calculation was performed for
   $n_\infty =0.15\; {\rm cm}^{-3},$ $T_\infty =1$keV, and $M = 3.6 \times 10^9 M_{Sun}.$
 The integration was from point $r =10 r_B.$
 The cooling through electron thermal conduction and radiative losses
   causes the atmosphere to be restructured qualitatively compared to the adiabatic case.
 More specifically, in the lower part of the atmosphere adjacent to the CFB,
   the temperature decreases greatly, while the density rises exponentially.

% ------- -- --- --- --- --- --- --- ------- Ris-04
\begin{figure}[t]
 \includegraphics[width=1\columnwidth]{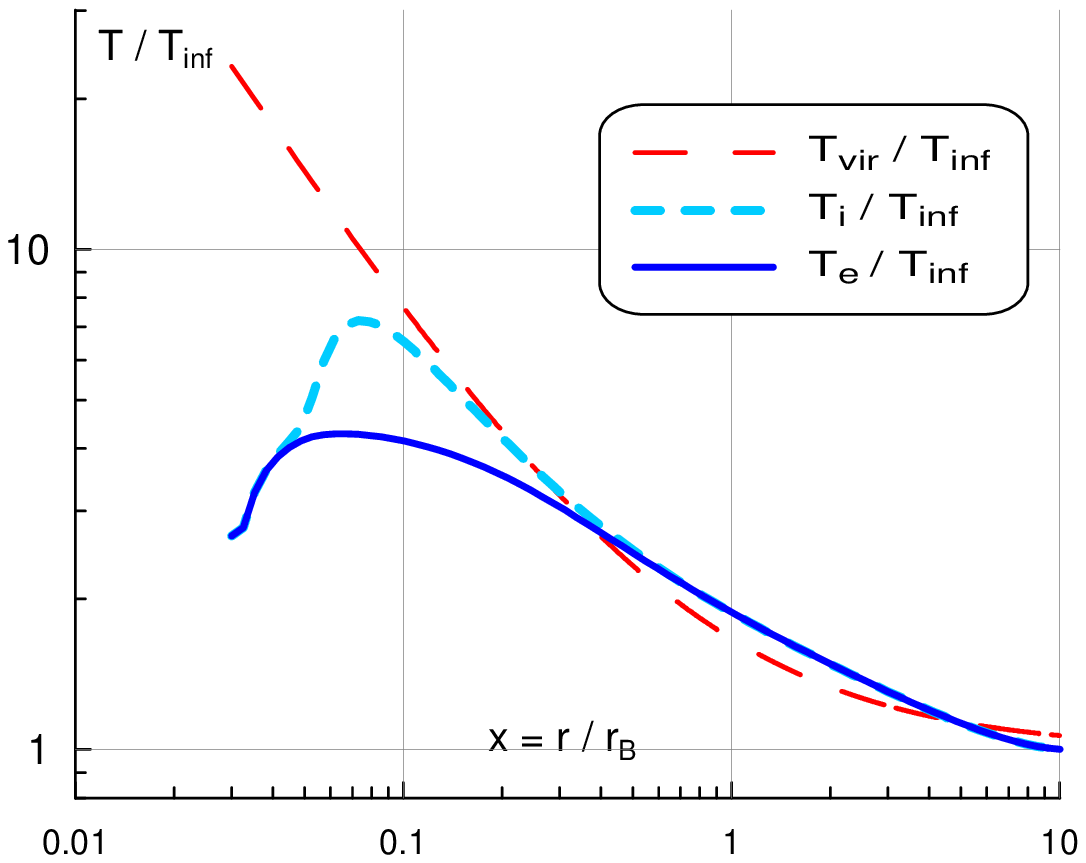}
 \caption{\label{fig:4}
 Distributions of the electron, $T_e,$ (solid curve) and ion, $T_i,$ (short dashes) temperatures
   in the accreting atmosphere above the CFB at a typical value of $x_c =0.03$
     that follows from turbulent estimates in Appendix 1.
 The parameter $x_c$ (1) characterizes the rotation velocity at the Bondi radius.
 The values of $T$ were normalized to the temperature at infinity.
 For comparison, the long dashes indicate the variation in virial temperature $T_{vir}.$
 As we see, the virial atmosphere is much hotter in the lower part.
      }
\end{figure}

% ------- -- --- --- --- --- --- --- ------- Ris-05
\begin{figure}[t]
 \includegraphics[width=1\columnwidth]{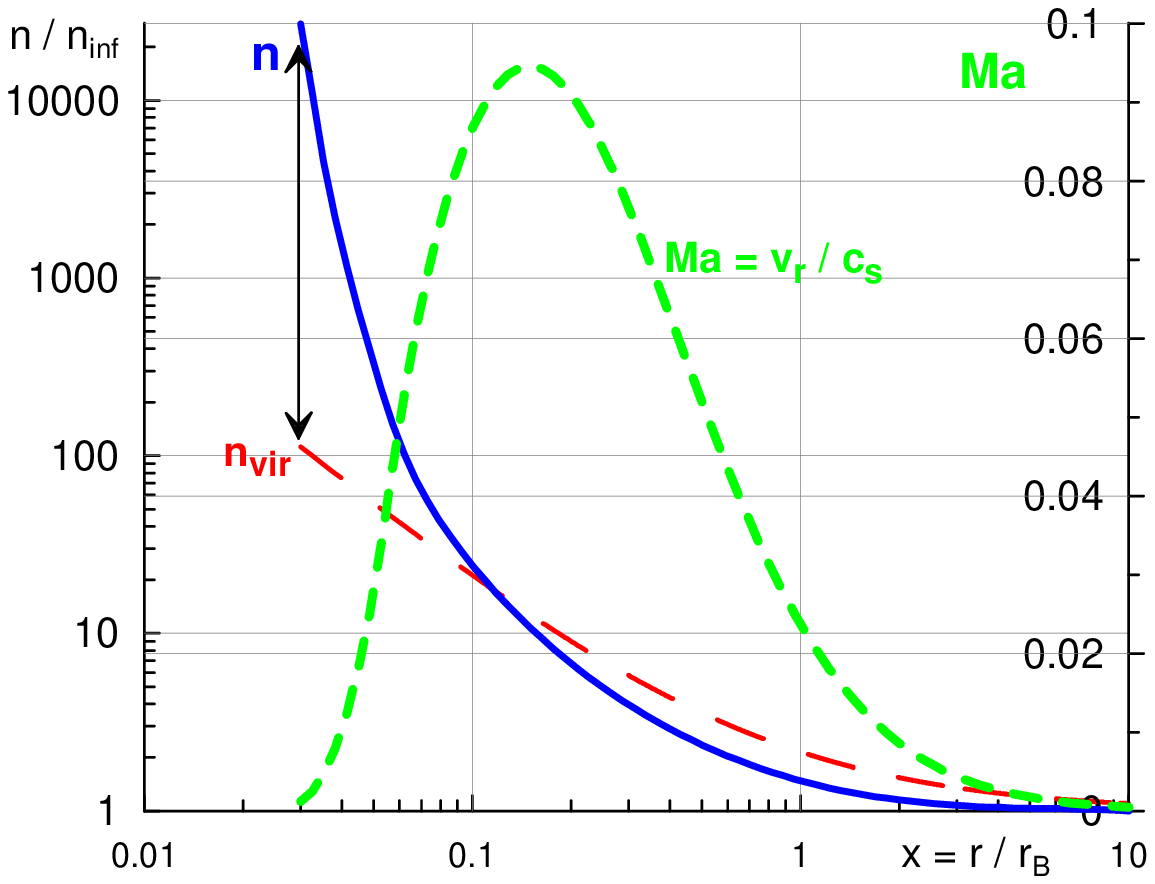}
 \caption{\label{fig:5}
 Sharp density rise in an atmosphere with cooling: cf. $n(r)$ (solid curve) for the solution of (14)-(18)
   with the virial distribution $n_{vir}(r)$ (the curve with long dashes).
 The arrow marks the enormous exponential density difference between $n(r)$ and $n_{vir}(r).$
 Due to the presence of CFB, the matter settling is subsonic -
   see the dependence of the Mach number on radius (the curve with short dashes).
 Case $x_c =0.03.$
      }
\end{figure}

 The dramatic increase in the "weight" of the atmosphere (the rise in $\rho,$ see Fig. 5)
   maintained in the gravity field through an approximately radial pressure gradient
     is associated with cooling.
 The accumulation of great masses roughly follows a hydrostatic balance with a decreased plasma temperature.
 A very dramatic rise in density makes the bremsstrahlung losses (10) efficient.
 Thus, the matter heated by gravitational contraction in Fig. 4 cools down in two stages.
 Initially, a condensation is formed through electron thermal conduction
   and, subsequently, the mechanism of radiative losses is switched on in a high-density plasma,
     which increases the density even more.

 The temperature inversion shown in Fig. 4 takes place due to heavy bremsstrahlung losses
   in the lower part of the atmosphere.
 The electron heat flux $q$ (7) changes its sign at the point of maximum of $T_e(r).$
 Below the maximum, the flux $q$ heats (!) the plasma, partially compensating for the bremsstrahlung losses.
 The boundary condition $T'_e(x_c)=0$ (17) is met in a small neighborhood of point $x_c,$
   which is not seen in the scale of Fig. 4.
 The intermediate region between the Bondi radius and the lower part of the atmosphere
   is thermally most stressed in flux $q.$
 The flux $q$ in this region is of the order of the limiting flux (9),
   while the electron-ion collision frequency (12)
     is insufficient to equalize the temperatures $T_e$ and $T_i.$

 In the intermediate region (narrow throat), the plasma is non-isothermal, $T_e < T_i$ (see Fig. 4).
 Above this region, isothermality is maintained by an expansion of the area of the sphere $4 \pi r^2$
   through which the electron heat removal goes.
 In this case, the heat flux $q$ per unit area of the sphere decreases.
 Below the throat, isothermality in Fig. 4 is restored by an exponential density rise.
 The point is that the limiting flux (9) is proportional to the density.

 The atmosphere is found to be essentially subsonic (see Fig. 5).
 The reduction in Mach number Ma at infinity is caused by an expansion of the sphere $4 \pi r^2.$
 In the lower part of the atmosphere, Ma decreases rapidly due to a sharp density rise
   and, accordingly, a rapid decrease in radial settling velocity according to integral (13).
 The maximum of Ma is associated with the intermediate region.
 Since the radial velocities $v_r$ are low compared to the speed of sound,
   the accretion rate $\dot M$ is also found to be low compared to the limiting one (13)
     which corresponds to the critical Bondi flow with $\gamma=5/3.$
 The dependence of the dimensionless $\dot m$ (13) on the rotation velocity at the Bondi radius
   for typical rotation velocities
     is shown in Fig. 6.
 As we see, a significant reduction in accretion rate compared to the limiting possible one (13)
   is associated with the CFB.

% ------- -- --- --- --- --- --- --- ------- Ris-06
\begin{figure}[t]
 \includegraphics[width=1\columnwidth]{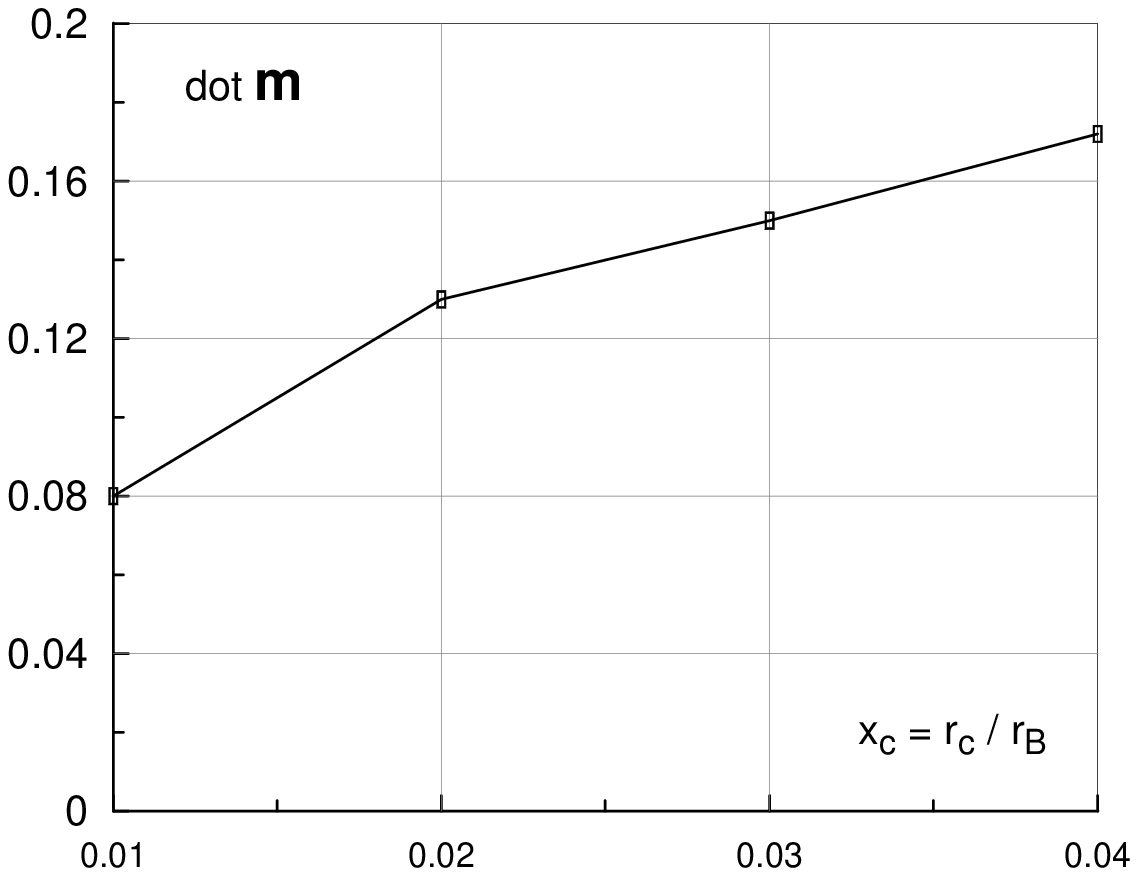}
 \caption{\label{fig:6}
 Influence of angular momentum on the accretion rate for typical (see Appendix 1) rotation velocities
   of the order of 100 km s$\!^{-1}$ at $r_B.$
 The remaining parameters are the same as those in Figs. 4 and 5.
      }
\end{figure}

 In conclusion, note that we considered the influence of electron thermal conduction
   on the structure of an accretion flow with gas rotation in the case of a SMBH.
 We showed that the combined action of thermal conduction and radiative cooling
   gives rise to a very massive thin cold torus and two disks, inner and outer.
 Due to heavy radiative losses, the inner disk near the torus and the outer disk
   all the way from the torus to the Bondi radius
     are found to be dense and thin.
 At smaller radii, the inner disk under the torus is restructured into a non-radiative ADAF.
 This is related to the conditions on the plasma density and temperature at the Bondi radius,
   under which a deficit of matter takes place.
 Because of this deficit, the emissivity of the inner disk at small radii under the torus
   is insufficient to maintain the regime of a standard disk by Shakura and Sunyaev (1973) (whence ADAF).
 The angular momentum of the accreting matter is removed over the outer disk.
 The outward transport of part of the matter beyond the Bondi radius
   is associated with the mechanical removal of angular momentum.
 This matter is in Keplerian rotation.
 The possibility of outward transport of mass and angular momentum through the settling flow in the atmosphere
   is related to strong cooling and a high density of the matter being transported.

 One of authors (R.A. Sunyaev) thanks Scott Tremaine and Glenn van de Ven for discussing the problem
   concerning the "spin" of galaxy M87.
 Authors are very grateful to V. Astakhov for his careful translation.

\vspace{1cm}

\centerline{REFERENCES}

 1. M. Abramowicz, M. Jaroszynski, and M. Sikora,
    Astron. Astrophys. {\bf 63,} 221 (1978).

 2. C.A.Back, D.H.Kalantar, R.L.Kauffman, et al.,
    Phys. Rev. Lett. {\bf 77,} 4350 (1996).

 3. S.A. Balbus and J.F. Hawley,
    Astrophys. J. {\bf 376,} 214 (1991).

 4. V.S. Beskin and L.M. Malyshkin,
    Astron. Lett. {\bf 22,} 475 (1996).

 5. R.D. Blandford and R.L. Znajek,
    Mon. Not. R. Astron. Soc. {\bf 179,} 433 (1977).

 6. E. Churazov, R. Sunyaev, W. Forman, et al.,
    Mon. Not. R. Astron. Soc. {\bf 332,} 729 (2002).

 7. T. Di Matteo, S.W. Allen, A.C. Fabian, et al.,
    Astrophys. J. {\bf 582,} 133 (2003).

 8. E. Emsellem, M. Cappellari, D. Krajnovic, et al.,
    Mon. Not. R. Astron. Soc. {\bf 379,} 401 (2007).

 9. A.C. Fabian and M.J. Rees,
    Mon. Not. R. Astron. Soc. {\bf 277,} L55 (1995).

 10. H.C. Ford, R.J. Harms, Z.I. Tsvetanov, et al.,
     Astrophys. J. {\bf 435,} L27 (1994).

 11. W. Forman, C. Jones, E. Churazov, et al.,
     Astrophys. J. {\bf 665,} 1057 (2007).

 12. K. Gebhardt and J. Thomas, arXiv: 0906.1492v2 [astro-ph.CO] (2009).

 13. R.J. Harms, H.C. Ford, Z.I. Tsvetanov, et al.,
     Astrophys. J. {\bf 435,} L35 (1994).

 14. S. Kato, J. Fukue, and S. Mineshige, Black-Hole Accretion Disks: Towards a New Paradigm
     (Kyoto Univ., Kyoto, 2008).

 15. P.I. Kolykhalov and R.A. Sunyaev, Pis'ma Astron. Zh. {\bf 6,} 680 (1980)
     [Sov. Astron. Lett. {\bf 6,} 357 (1980)].

 16. M. Kozlowski, M. Jaroszynski, and M. Abramowicz,
     Astron. Astrophys. {\bf 63,} 209 (1978).

 17. E.V. Levich and R.A. Syunyaev, Astron. Zh. {\bf 48,} 461 (1971) [Sov. Astron. {\bf 15,} 363 (1971)].

 18. D.N. Limber, Astrophys. J. {\bf 140,} 1391L (1964).

 19. F. Macchetto, A. Marconi, D.J. Axon, et al., Astrophys. J. {\bf 489,} 579 (1997).

 20. M. Miyoshi, J. Moran, J. Herrnstein, et al., Nature {\bf 373,} 127 (1995).

 21. R. Narayan and I. Yi, Astrophys. J. {\bf 428,} L13 (1994).

 22. F.N. Owen, J.A. Eilek, and N.E. Kassim, Astrophys. J. {\bf 543,} 611 (2000).

 23. B. Paczynski and P. J. Witta, Astron. Astrophys. {\bf 88,} 23 (1980).

 24. V.F. Schwartzman, Astron. Zh. {\bf 48,} 479 (1971) [Sov. Astron. {\bf 15,} 377 (1971)].

 25. N.I. Shakura and R.A. Sunyaev, Astron. Astrophys. {\bf 24,} 337 (1973).

 26. L. Spitzer, Phys. of Fully Ionized Gases, 2nd ed. (Intersci., New York, 1962), p. 170.

 27. V. Springel, T. Di Matteo, and L. Hernquist, Mon. Not. R. Astron. Soc. {\bf 361,} 776 (2005).

 28. K.P. Stanyukovich, Unsteady Motion of Continuous Media (Pergamon, New York, 1960; Nauka, Moscow, 1971).

 29. E.P. Velikhov, Zh. Eksp. Teor. Fiz. {\bf 36,} 1398 (1959) [Sov. Phys. JETP {\bf 9,} 995 (1959)].

 30. A.J. Young, A.S. Wilson, and C.G.Mundell, Astrophys. J. {\bf 579,} 560 (2002).

 31. N.L. Zakamska and R. Narayan, Astrophys. J. {\bf 582,} 162 (2003).

\end{document}